\documentclass[12pt,preprint]{emulateapj}

\begin{document}

\title{Multiplicity and Optical Excess Across the Substellar Boundary in Taurus}

\author{Adam L. Kraus (alk@astro.caltech.edu), Russel J. White (rjw@astro.caltech.edu), and Lynne A. Hillenbrand (lah@astro.caltech.edu)}
\affil{California Institute of Technology, Department of Astrophysics, MC 105-24, Pasadena, CA 91125}

\begin{abstract}

We present the results of a high-resolution imaging survey of 22 brown dwarfs and
very low mass stars in the nearby ($\sim$145 pc) young ($\sim$1-2 Myr) low-density
star-forming region Taurus-Auriga.  We obtained images with the Advanced Camera for
Surveys/High Resolution Channel on HST through the $F555W$ \,($V$), $F775W$
\,($i'$), and $F850LP$ \,($z'$) filters. This survey confirmed the binarity of
MHO-Tau-8 and discovered a new candidate binary system, V410-Xray3, resulting in a
binary fraction of $9\pm5\%$ at separations $>$4 AU. Both binary systems are tight
($<$10 AU) and they possess mass ratios of 0.75 and 0.46, respectively.  The binary
frequency and separations are consistent with low-mass binary properties in the
field, but the mass ratio of V410-Xray3 is among the lowest known. We find that the
binary frequency is higher for very low mass stars and high-mass brown dwarfs than
for lower-mass brown dwarfs, implying either a decline in frequency or a shift to
smaller separations for the lowest mass binaries. Combining these results with
multiplicity statistics for higher-mass Taurus members suggests a gradual decline
in binary frequency and separation toward low masses. The implication is that the
distinct binary properties of very low-mass systems are set during formation and
that the formation process is similar to the process which creates higher-mass
stellar binaries, but occurs on a smaller scale. We combine the survey detection
limits with models for planetary-mass objects to show that there are no planets or
very low-mass brown dwarfs with mass $>3 M_J$ at projected separation $>$40
AU orbiting any of the Taurus members in our sample, implying that planetary-mass
companions at wide separations are rare. Finally, based on fits to the optical and
near-infrared spectral energy distributions, we identify several BDs with
significant ($\ga$1 mag) V-band excesses. The excesses appear to be correlated with
signatures of accretion, and if attributed to accretion luminosity, may imply mass 
accretion rates several orders of magnitude above those inferred from line-profile 
analyses.

\end{abstract}


\section{Introduction}

Brown dwarfs (BDs) are objects with masses between those of stars and planets,
insufficient to maintain stable fusion reactions in their cores. BDs comprise a
significant fraction of the total (sub)stellar content of the galaxy and are
among our nearest neighbors (Reid et al. 2004); in the past decade, field
surveys have discovered several hundreds of BDs in the solar neighborhood (e.g.
Delfosse et al. 1997; Kirkpatrick et al. 1999; Leggett et al.  2000).  Soon
after BDs were discovered, it was found that many, like stars, are members of
binary systems. However, the properties of binary systems near and below the
substellar boundary ($M_{primary}<0.2 M_{\sun}$) appear to be fundamentally
different from those of higher-mass stars ($0.3<M_{primary}<1.0$$M_{\sun}$).  
Multiplicity surveys of field T dwarfs \citep{burg03}, L dwarfs
\citep{K99,C03,bouy03,G03}, and late M dwarfs \citep{sieg05} have found lower
binary frequencies ($\sim$15\% vs 40-55\%) and smaller binary separations ($<$20
AU vs $<1000$ AU) than for field stars (Duquennoy \& Mayor 1991; Fischer \&
Marcy 1992; Halbwachs et al. 2003).

These results demonstrate that field binary properties depend on mass.  
Unfortunately, binary frequencies for field stars are only reported for broad
mass ranges, so they do not place strong constraints on the functional form of
this dependence. Various groups interpret the transition in binary properties as
either a sharp break near the stellar/substellar boundary \citep{krou03,C03} or
a smooth mass dependence \citep{luh04b}. Also, since field BD systems are
typically old and possess lower binding energies than stellar binaries of equal
separation, the results could be biased by the dynamical disruption of wide,
low-mass systems. To begin testing this possibility, we performed a small
multiplicity survey of low-mass stars and BDs in the nearby young OB Association
Upper Scorpius (Kraus et al. 2005) and T Association Taurus (this work). In
Upper Sco, we found several young binary systems, but the binary frequency,
separations, and mass ratios were consistent with the field and somewhat
($\sim$2$\sigma$) inconsistent with higher-mass members of Upper Sco (Kohler et
al. 2000). This suggests that dynamical evolution after the T Tauri stage
probably does not produce the unique binary parameters of BDs;  instead, the
implication is that the mechanism by which binaries form depends on mass.

Recent efforts to model low mass binary formation have typically assumed that a
cluster of 5-10 protostellar embryos form from a single fragmenting cloud core
(e.g. Kroupa 1995; Sterzik \& Durisen 1998; Kroupa \& Bouvier 2003; Kroupa et al.
2003; Delgado-Donate et al. 2003; Hubber et al. 2005); these embryos would then
undergo dynamical evolution to form single stars and stable multiple systems. 
However, the frequency of multiple stellar systems ($M_{primary}>0.3 M_{\sun}$)
in the field (35-57\%;  Duquennoy \& Mayor 1992; Reid \& Gizis et al. 1997) and
in young associations (50-80\%; Kohler et al. 2000; White et al. 2006) has
been interpreted by Goodwin \& Kroupa (2005) to mean that the collapse and
fragmentation of a cloud core produces only 2 or 3 stars. Larger systems would
eject more single stars and tight binaries than are observed.  Reipurth \&
Clarke (2001) have suggested that evolution to a dynamically stable state could
occur early in the formation process; the ejected embryos would then cease
accretion and become BDs.

The ejection process would preferentially disrupt wide BD binary systems, causing
the deficit of wide systems seen in low-mass field binaries. However, simulations
by Bate et al.  (2003) find that the corresponding binary frequencies and
separations ($<$5\% and $<$10 AU) are too low to be consistent with the field.  
Also, some wide very-low-mass binaries have recently been discovered in the field
(Gizis et al. 2000; Phan-Bao et al. 2005; Billeres et al. 2005). These systems are
very weakly bound and most likely would not survive the ejection process, though
recent simulations by Bate et al. (2005) suggest that they could form via
simultaneous ejection of previously-unbound objects. Finally, some models predict
that ejection would alter other properties of BDs (spatial and velocity dispersion,
disk lifetime, and accretion frequency). The preponderance of observations show
that these properties are similar in the stellar and substellar regimes
\citep{WB03,luh04b,wh04,moh05}; the strong similarity between the two regimes
suggests that brown dwarfs for in a manner similar to stars and thus that BD
binaries form like stellar binaries, though possibly on a smaller scale. The
implication is that all binaries share a common formation mechanism, fragmentation
of a single collapsing cloud core, and that this mechanism features a mass
dependence that remains unexplained by theoretical models.

Multiplicity surveys of the field and of nearby stellar populations (e.g. Kohler et
al. 2000; Luhman et al. 2005a; White et al. 2006) have placed some constraints on
the form of this mass dependence. In particular, White et al. (2006) studied the
mass dependence of multiplicity in a speckle interferometry survey of the nearby
($\sim$145 pc) young ($\sim$1-2 Myr) T Association Taurus-Auriga. This
surveyincluded objects from 1.5 $M_\sun$ to the substellar boundary and found that
the separation distribution and mass ratio distribution functions were mass
dependent. Their results for the binary frequency were inconclusive, but suggested
a possible slow decline with mass.  Better statistics will be required for very low
mass binaries in order to confirm this trend. In this paper, we present the results
of an complementary imaging multiplicity survey of very low mass stars and brown
dwarfs to near the planetary mass regime in Taurus-Auriga.

\section{Observations and Data Reduction}

\subsection{Sample Selection}

The Taurus-Auriga association has been the target of many recent wide-field surveys
to detect new low-mass members (e.g. Briceno et al. 1998; Martin et al.  2001;
Briceno et al. 2002; Luhman et al. 2003a). These surveys identified candidate
members based on their location on an optical or near-infrared color-magnitude
diagram, and membership was then confirmed spectroscopically via the detection of
lithium absorption, excess $H\alpha$ emission, or low surface gravity, all of which
are indicators of youth. We selected all 18 confirmed Taurus members with spectral
type later than M5.5 discovered by these surveys.  Three additional targets
(V410-Xray3, V410-Anon13, and GM Tau) are previously known Taurus members which
have also been confirmed to possess spectral types in this range (Strom \& Strom
1994; White \& Basri 2003). Our final target (LH0419+15)  was chosen from a survey
for Hyades members by Reid \& Hawley; it is the only member of their survey with a
spectroscopic detection of lithium, and they classify it as a likely Taurus member
with spectral type M7 based on its apparent youth and distance.  We list these 22
targets and their discovery or confirmation references in Table 1; this was a
complete list of known Taurus members with spectral types later than M5.5 at the
time the observations were proposed (January 2003). Since our targets include both
very low mass stars and brown dwarfs, we hereafter refer to them as very low mass
objects, or VLMOs.

\subsection{Observations}

Our images were obtained with the Advanced Camera for Surveys/High Resolution
Camera on the Hubble Space Telescope, which has a field of view of 26x29\arcsec and
distortion-corrected pixel size of 27 mas pix$^{-1}$. In Table 2, we summarize the
exposure time and epoch of observation for each target (Program ID: 9853).
Observations were made between September 2003 and January 2004 with the filters
F555W (V), F775W (i'), and F850LP (z') at two dither positions near the center of
the detector and with two exposures per position. Total integration times were 510,
300, and 200 seconds, respectively. The F555W exposure times for the brightest
objects were reduced to 350 seconds to allow for additional short exposures in
F775W and F850LP, which were close to the saturation limit in the full-length
exposures and were saturated in the case of MHO-Tau-5 and MHO-Tau-8.  Saturation
was permitted in the full-length images to allow for comparable sensitivities to
faint companions at wide separations. We chose the V band to maximize angular
resolution (diffraction limit $\theta_{res,V}=58$ mas)  and the i' and z' bands to
maximize sensitivity to very low mass companions.

The raw images were calibrated and distortion-corrected by the CALACS 
pipeline during on-the-fly-reprocessing \citep{mack03}. Some cosmic rays 
remained, but their morphologies were substantially different from stellar 
PSFs, so they were easily identified by visual inspection. 

\begin{deluxetable*}{llllll}
\tabletypesize{\scriptsize}
\tablewidth{0pt}
\tablecaption{Observations\label{tbl1}}
\tablehead{\colhead{Target} & \colhead{Date\tablenotemark{a}} &
\multicolumn{3}{c}{Exposure Times (s)} & \colhead{Discovery}
\\
\colhead{} & \colhead{} & \colhead{F555W}&\colhead{F775W}&\colhead{F850LP}
}
\startdata
CFHT-Tau-1&2904.3&510&300&200&Martin et al. (2001)\\
CFHT-Tau-2&2904.3&510&300&200&Martin et al. (2001)\\
CFHT-Tau-3&2901.5&510&300&200&Martin et al. (2001)\\
CFHT-Tau-4&2857.5&510&300&200&Martin et al. (2001)\\
KPNO-Tau-1&2881.2&510&300&200&Briceno et al. (2002)\\
KPNO-Tau-2&2903.3&510&300&200&Briceno et al. (2002)\\
KPNO-Tau-3&2901.4&510&300&200&Briceno et al. (2002)\\
KPNO-Tau-4&2904.4&510&300&200&Briceno et al. (2002)\\
KPNO-Tau-5&2884.2&350&300/30&200/20&Briceno et al. (2002)\\
KPNO-Tau-6&3054.5&510&300&200&Briceno et al. (2002)\\
KPNO-Tau-7&2901.6&510&300&200&Briceno et al. (2002)\\
KPNO-Tau-8&2850.5&350&300/30&200/20&Briceno et al. (2002)\\
KPNO-Tau-9&3025.8&510&300&200&Briceno et al. (2002)\\
KPNO-Tau-12&3028.7&510&300&200&Luhman et al. (2003a)\\
KPNO-Tau-14&2897.5&350&300/30&200/20&Luhman et al. (2003a)\\
MHO-Tau-4&2893.5&350&300/30&200/20&Briceno et al. (1998)\\
MHO-Tau-5&3023.0&350&300/30&200/20&Briceno et al. (1998)\\\
MHO-Tau-8&3028.5&350&300/30&200/20&Briceno et al. (1998)\\\
LH 0419+15&2901.0&510&300&200&Reid \& Hawley (1999)\\
V410-Xray3&3029.4&350&300/30&200/20&Strom \& Strom (1994)\\
V410 Anon13&2903.4&510&300&200&Strom \& Strom (1994)\\\
GM Tau&2850.6&350&300/30&200/20&White \& Basri (2002)\\
\enddata
\tablenotetext{a}{Observation Date: JD minus 2450000.}
\end{deluxetable*}

\subsection{Data Reduction}

Potential point sources were identified with the IRAF task DAOPHOT/DAOFIND, and
we measured aperture photometry and point-spread function (PSF) fitting
photometry for all objects in each field using the DAOPHOT tasks PHOT and
ALLSTAR. We describe these data reduction procedures in more detail in our
companion paper on VLMO multiplicity in Upper Sco (Kraus et al. 2005) and the
DAOPHOT package is described by Stetson et al. (1987). We report aperture
photometry for all isolated objects and PSF photometry for all close binaries.
PSF magnitudes were corrected to match aperture magnitudes based on results for
the twenty apparently isolated Taurus members in our sample (Section 3.1). A
preliminary PSF for each filter was constructed from the 20 well-sampled VLMOs
that appeared isolated under visual inspection. Since all of the targets were
located near the center of the chip and have similar temperatures, image
distortion and target color should not be important. There was some variation in
the PSF FWHM from target to target ($\pm5\%$), which we attribute to small
orbit-to-orbit changes in focus. We investigated this by dividing our sample
into two groups, based on whether the target PSF appeared to be narrower or
wider than the average PSF, and constructing new average PSFs for each group. We
then re-ran PSF photometry, but the modest decrease in the residuals did not
reveal any companions which were not previously identified. Since it is not
possible to determine which PSF is appropriate for blended binaries, we proceed
using only the average PSF for the entire group. This choice could lead to
systematic errors in the calculation of binary properties; we discuss these
errors in more detail in Section 3.4.

Transformations to ground-based magnitudes ($V$, SDSS $i'$, and SDSS 
$z'$) were described in detail in Kraus et al. (2005). We found constant 
corrections that do not depend significantly on temperature or surface gravity:
$m_{555}-m_V=-0.16\pm0.03$, $m_{775}-m_{i'}=+0.07\pm0.03$, and 
$m_{850}-m_{z'}=+0.03\pm0.03$. The transformed magnitudes and 
uncertainties are listed in Table 2. The statistical uncertainties 
correspond to either the photon noise (for aperture photometry; typically 
$<$0.01 magnitudes) or the goodness of fit (for PSF-fitting photometry; 
0.01-0.06 magnitudes for objects with significant fits). Systematic 
uncertainties in the magnitude transformations and aperture corrections 
are $\sim$0.03 magnitudes. The photometry calculated from short exposures was 
consistent with that from long exposures, so we report only the long exposures.

\begin{deluxetable*}{lllllll}
\tabletypesize{\scriptsize}
\tablewidth{0pt}
\tablecaption{Photometry of Very Low Mass Objects in Taurus\label{tbl2}}
\tablehead{\colhead{Name}&\colhead{$V$\tablenotemark{a}} 
&\colhead{$i'$\tablenotemark{a}}&\colhead{$z'$\tablenotemark{a}} 
&\colhead{$J$\tablenotemark{b}} & \colhead{$H$\tablenotemark{b}} 
&\colhead{$K$\tablenotemark{b}}
}
\startdata
CFHT-Tau-1&23.427$\pm$0.026&18.612$\pm$0.003&16.612$\pm$0.002&13.740$\pm$0.024&12.537$\pm$0.023&11.849$\pm$0.015\\
CFHT-Tau-2&22.520$\pm$0.014&18.112$\pm$0.002&16.307$\pm$0.002&13.754$\pm$0.021&12.762$\pm$0.021&12.169$\pm$0.017\\
CFHT-Tau-3&21.654$\pm$0.009&17.970$\pm$0.002&16.119$\pm$0.001&13.724$\pm$0.023&12.861$\pm$0.023&12.367$\pm$0.023\\
CFHT-Tau-4&21.556$\pm$0.008&16.920$\pm$0.001&14.951$\pm$0.001&12.168$\pm$0.020&11.008$\pm$0.019&10.332$\pm$0.016\\
KPNO-Tau-1&24.063$\pm$0.044&19.572$\pm$0.004&17.596$\pm$0.003&15.101$\pm$0.038&14.247$\pm$0.037&13.772$\pm$0.034\\
KPNO-Tau-2&21.376$\pm$0.007&17.670$\pm$0.002&16.074$\pm$0.001&13.925$\pm$0.022&13.241$\pm$0.027&12.753$\pm$0.020\\
KPNO-Tau-3&20.239$\pm$0.004&16.959$\pm$0.001&15.484$\pm$0.001&13.323$\pm$0.019&12.501$\pm$0.021&12.079$\pm$0.019\\
KPNO-Tau-4&24.722$\pm$0.071&20.072$\pm$0.006&17.897$\pm$0.004&14.997$\pm$0.033&14.025$\pm$0.037&13.281$\pm$0.031\\
KPNO-Tau-5&19.690$\pm$0.004&16.226$\pm$0.001&14.706$\pm$0.001&12.640$\pm$0.020&11.918$\pm$0.022&11.536$\pm$0.016\\
KPNO-Tau-6&22.292$\pm$0.012&19.097$\pm$0.003&17.301$\pm$0.003&14.995$\pm$0.028&14.197$\pm$0.038&13.689$\pm$0.036\\
KPNO-Tau-7&22.068$\pm$0.011&18.358$\pm$0.002&16.661$\pm$0.002&14.521$\pm$0.030&13.828$\pm$0.026&13.272$\pm$0.032\\
KPNO-Tau-8&19.261$\pm$0.003&16.147$\pm$0.001&14.863$\pm$0.001&12.946$\pm$0.018&12.367$\pm$0.019&11.985$\pm$0.020\\
KPNO-Tau-9&24.918$\pm$0.084&20.035$\pm$0.006&18.056$\pm$0.004&15.497$\pm$0.042&14.660$\pm$0.039&14.185$\pm$0.053\\
KPNO-Tau-12&23.228$\pm$0.023&20.781$\pm$0.009&18.998$\pm$0.006&16.305$\pm$0.085&15.483$\pm$0.096&14.927$\pm$0.092\\
KPNO-Tau-14&20.735$\pm$0.007&16.297$\pm$0.001&14.502$\pm$0.001&11.907$\pm$0.019&10.805$\pm$0.021&10.269$\pm$0.018\\
MHO-Tau-4&18.678$\pm$0.002&15.246$\pm$0.001&13.732$\pm$0.001&11.653$\pm$0.028&10.923$\pm$0.032&10.567$\pm$0.020\\
MHO-Tau-5&17.595$\pm$0.001&14.489$\pm$0.001&13.114$\pm$0.001&11.070$\pm$0.023&10.390$\pm$0.029&10.063$\pm$0.020\\
MHO-Tau-8&17.951$\pm$0.002&14.453$\pm$0.001&12.976$\pm$0.001&10.862$\pm$0.018&10.140$\pm$0.020&9.726$\pm$0.016\\
MHO-Tau-8 A&18.241$\pm$0.030&14.911$\pm$0.033&13.556$\pm$0.021&&&\\
MHO-Tau-8 B&19.525$\pm$0.048&15.611$\pm$0.060&13.934$\pm$0.013&&&\\
LH 0419+15&21.835$\pm$0.010&18.028$\pm$0.002&16.488$\pm$0.002&14.364$\pm$0.029&13.549$\pm$0.027&13.079$\pm$0.035\\
V410 Xray-3&18.319$\pm$0.002&15.046$\pm$0.001&13.629$\pm$0.001&11.544$\pm$0.018&10.817$\pm$0.021&10.446$\pm$0.017\\
V410 Xray-3 A&...&15.046$\pm$0.014&13.812$\pm$0.011&&&\\
V410 Xray-3 B&...&20.032$\pm$0.446&15.658$\pm$0.028&&&\\
V410 Anon-13&22.175$\pm$0.012&17.665$\pm$0.002&15.782$\pm$0.001&12.954$\pm$0.019&11.659$\pm$0.020&10.958$\pm$0.015\\
GM Tau&17.577$\pm$0.001&15.169$\pm$0.001&13.908$\pm$0.001&12.804$\pm$0.019&11.586$\pm$0.017&10.632$\pm$0.016\\
\enddata
\tablenotetext{a}{Uncertainties are statistical only; systematic uncertainties 
due to aperture corrections and conversion to standard systems are 
$\sim$0.03 magnitudes.}
\tablenotetext{b}{Near-infrared photometry is taken from the Two Micron 
All Sky Survey (Cutri et al. 2003). We quote total system magnitudes for the two 
binary systems.}
\end{deluxetable*}

\section{Results}

\subsection{VLMO Binaries}

The present survey found no fully resolved low-mass binaries. 
However, as we discuss in Kraus et al. (2005), one limitation in 
ALLSTAR-based PSF photometry is that binaries with very close 
($\la$$\lambda$$/D$) separations are often not identified, 
even when their combined PSF is elongated at a high confidence level. 
DAOFIND, the task which identifies potential objects in the images, 
identifies point sources based only on the presence of a distinct peak. Thus, 
automated photometry will be biased against the detection of very close 
binaries. This limitation can be overcome for known or suspected binaries 
by manually adding a second point source in approximately the correct 
location and letting ALLSTAR recenter it to optimize the fit. We have 
already used this method to recover photometry for the candidate binary 
system USco-109 in Kraus et al. (2005).

In Figures 1-3, we illustrate this technique with contour plots of the known VLMO
binary MHO-Tau-8 (discovered by White et al. (2006) and independently rediscovered
here), the new candidate VLMO binary V410-Xray3, and the apparently single VLMO
MHO-Tau-4 in the F555W, F775W, and F850LP filters. Since the long exposures in i'
and z' for MHO-Tau-8 were saturated, we show only the short exposures. Neither of
the two candidate systems is obviously resolved and neither was consistently
reported as a double-source by DAOFIND, but the PSF for MHO-Tau-8 is slightly
elongated along the y-axis and that of V410-Xray3 is slightly elongated along the
x-axis relative to MHO-Tau-4.

In Figure 1, we present contour plots of MHO-Tau-8 in each filter. The first
column shows the original image, the second column shows the result from fitting
with one point source, and the last column shows the result from fitting with two
point sources. The maximum and minimum pixel values are also given to allow
quantitative comparison of the residuals to the original images. The common
position angle of the residuals in the single-source fit seems to imply that this
elongation is a real effect, and not simply noise. Normal levels of jitter were
reported in the observation log, so this is also unlikely to be a systematic
effect. Unfortunately, there were no other bright objects in the field to serve as
PSF references. However, as we summarize in Table 3, the system properties in each
filter are consistent with previous observations and expected orbital motion
(Section 5.3). Since MHO-Tau-8 was independently identified as a binary system by
White et al. (2006), we regard it as a confirmed discovery.

In Figure 2, we present similar contour plots of V410-Xray3, showing both 
the long and short exposures in i' and z'. The residuals from the 
single-source fit are roughly aligned and the jitter levels were normal, 
as in the case of MHO-Tau-8, but the separations of the residuals are 
marginally lower, implying a smaller separation or larger flux ratio. 
ALLSTAR was unable to fit two point sources in the V image and the fit for 
the secondary in the i' images was not statistically significant, but the 
fit for the z' images appears consistent and statistically significant in 
both the short and long exposures. The number statistics for the i' and z' 
images are similar and the z' filter has lower resolution, therefore the 
superior fit for the z' images suggests that a possible companion may be 
much cooler and redder. We summarize the system parameters as calculated 
from each image in Table 3, though the results for the i' images should be 
used with caution. 

The significant residuals in the double-source fit and the scatter in 
system properties suggests that even if V410-Xray3 is a binary, the 
measured parameters are not very reliable. Since there are no background 
stars for comparison and no other observations to support its 
multiplicity, we suggest its classification as a candidate binary. 
Followup observation to confirm its existence and properties should be a 
priority, since if confirmed, its small separation and corresponding short 
orbital period ($\sim$50 years for a circular orbit) could allow for a 
dynamical mass determination in less than a decade. Our subsequent 
analysis will consider V410-Xray3 as a candidate binary, but since its 
binarity has not been confirmed or disproved, our discussion will reflect 
both possibilities\footnote[1]{Observations conducted in February 2006 with the 
Keck-II telescope and Laser Guide Star Adaptive Optics have confirmed this 
candidate and will be reported in a future publication.}.

In Figure 3, we illustrate the typical results for single stars with contour plots
of MHO-Tau-4 and the residuals after fitting with a single point source. We could
not obtain a statistically significant fit for two point sources, suggesting that
there are no binary companions at separations $\ga$4 AU with mass ratio near unity.

One target VLMO, KPNO-Tau-14, was reported as a possible double-lined spectroscopic
binary by Mohanty et al. (2005). We did not detect any PSF elongation for this
target. Since the components of SB2s have similar brightness, the detection limits
we find in Section 3.3 imply its separation is less than our inner detection limit
($\sim$4 AU). Since it falls inside our survey limits, we will not consider it as a
binary in our discussion.

\begin{figure} 
\epsscale{1.00} 
\plotone{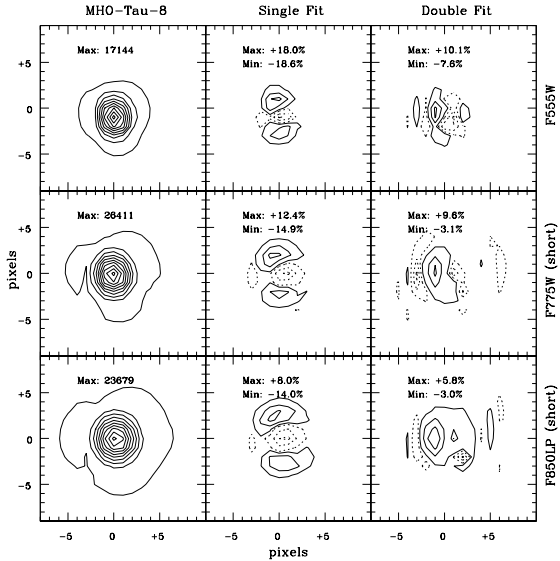} 
\caption{ 
Contour plots of MHO-Tau-8 for all three filters. The first column shows MHO-Tau-8,
the second column shows the residuals from fitting it with one source, and the last
column shows the residuals from fitting it with two sources. For residuals,
contours are drawn at the 90\%, 50\%, and 10\% levels of maximum (solid lines) and
minimum (dashed lines). The peak pixel value in each original image is shown; 
the positive and negative peaks of the residuals are reported as a percentage of 
the original peak value.
}
\end{figure}

\begin{figure}
\epsscale{1.00}
\plotone{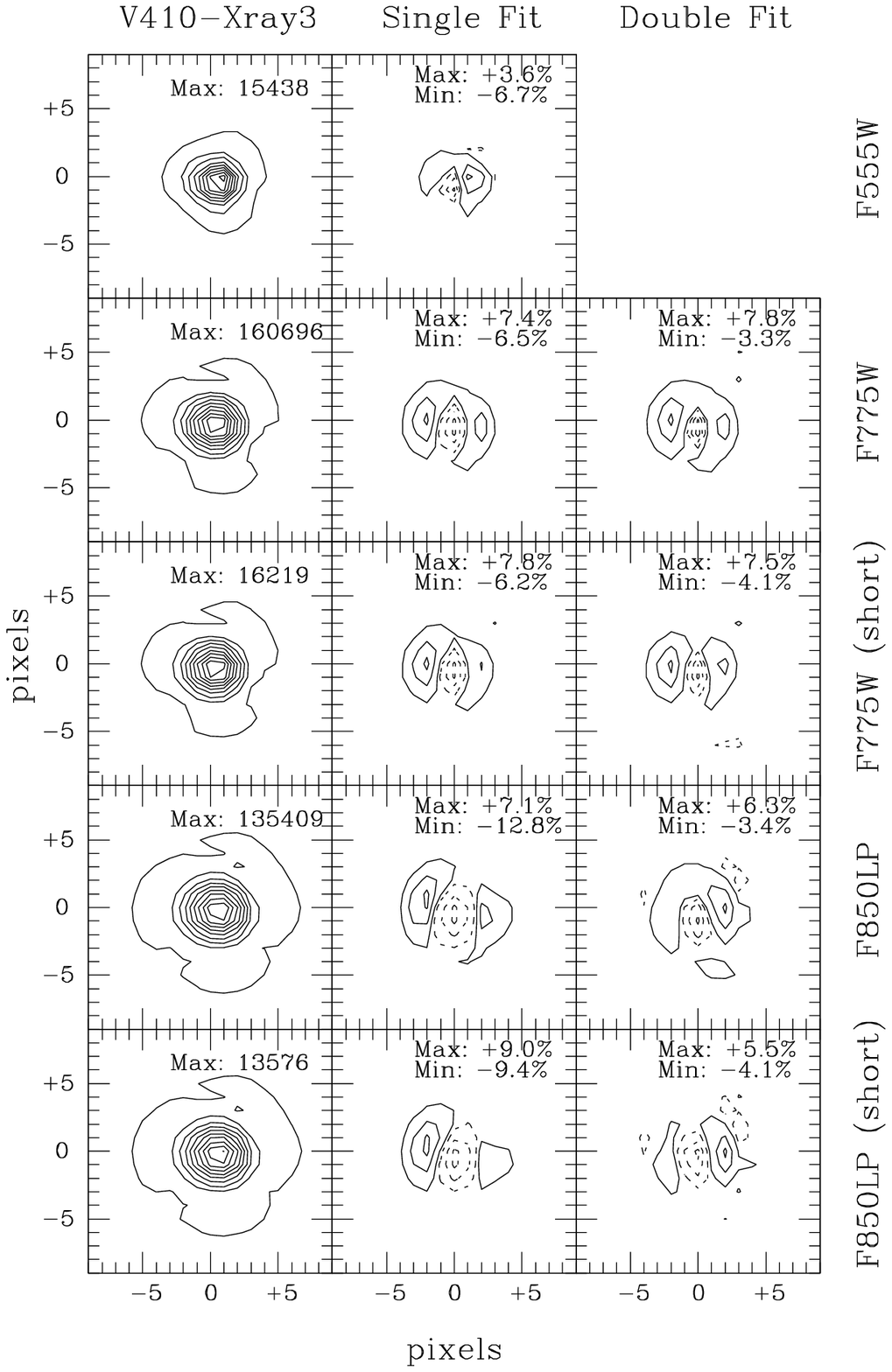}
\caption{
Contour plots of V410-Xray3 for all three filters. The first column shows 
V410-Xray3, the second column shows the residuals from fitting it with one 
source, and the last column shows the residuals from fitting it with two 
sources. For residuals, contours are drawn at the 90\%, 50\%, and 10\% 
levels of maximum (solid lines) and minimum (dashed lines). The peak pixel value 
in each original image is shown; the positive and negative peaks of the 
residuals are reported as a percentage of the original peak value.
}
\end{figure}

\begin{figure}
\epsscale{1.00}
\plotone{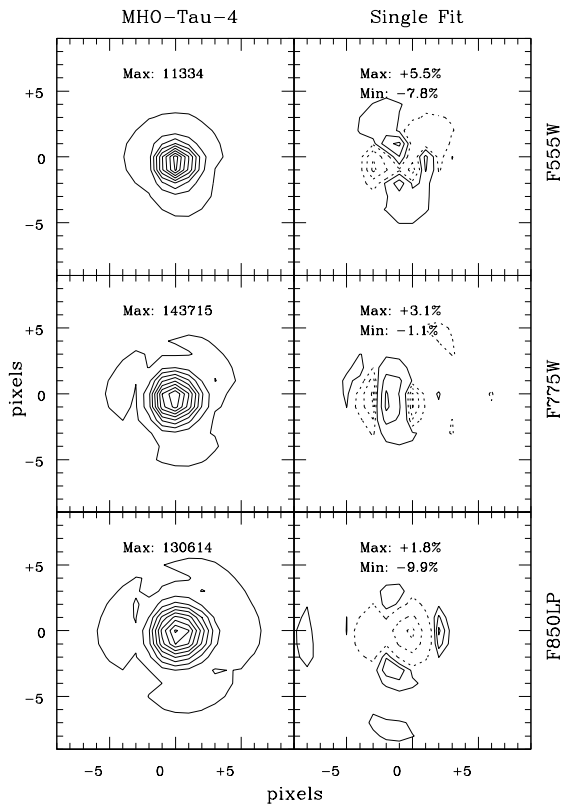}
\caption{
Contour plots of MHO-Tau-4 for all three filters. The first column shows
MHO-Tau-4 and the second column shows the residuals from fitting it with one
source. For residuals, contours are drawn at the 90\%, 50\%, and 10\% levels of
maximum (solid lines) and minimum (dashed lines). The peak pixel value in each
original image is shown;  the positive and negative peaks of the residuals are
reported as a percentage of the original peak value.
}
\end{figure}

\begin{deluxetable*}{llllllll}
\tabletypesize{\scriptsize}
\tablewidth{0pt}
\tablecaption{Binary Properties\label{tbl3}}
\tablehead{\colhead{Target} & \colhead{Date\tablenotemark{a}} &
\colhead{Separation} & \colhead{Position} & \colhead{Filter} &  
\colhead{Flux Ratio} & \colhead{$\Delta$$SpT$} & \colhead{$q$}
\\
\colhead{} & \colhead{} &\colhead{(mas)} & \colhead{Angle(deg)} 
& \colhead{} & \colhead{$\Delta$$m$ (mag)}
}
\startdata
MHO-Tau-8&1143\tablenotemark{b}&37&311&K&0.02&&\\
&3029&54&269&V&1.28&&\\
&3029&39&272&i'&0.70&&\\
&3029&40&268&z'&0.378&&\\
&$<$3029$>$\tablenotemark{c}&44$\pm$8&270$\pm$2&&&0.6$\pm$0.1&0.75$\pm$0.07\\
V410-Xray3&3029&56&192&i'(long)&5.0&&\\
&3029&45&205&z'&1.85&&\\
&3029&46&234&i'(short)&3.7&&\\
&3029&43&203&z'&1.55&&\\
&$<$3029$>\tablenotemark{c}$&44$\pm$2&204$\pm$2&&&1.7$\pm$0.2&0.47$\pm$0.07\\
\enddata
\tablenotetext{a}{Observation Date: JD minus 2450000.}
\tablenotetext{b}{White et al. (2006)}
\tablenotetext{c}{System values for MHO-Tau-8 are determined from the mean and 
standard deviation for all HST observations; values for V410-Xray3 are 
determined from the long and short $z'$ observations}
\end{deluxetable*}

\subsection{VLMOs and Background Stars}

In Figure 4, we present an $i'$ versus $i'-z'$ color magnitude diagram of the 20
apparently single VLMO targets (filled circles), the components of the two
candidate binary systems (filled circles with error bars), and all other objects
which were clearly resolved and detected at the $5\sigma$ level in both filters
(open circles). The candidate secondary component of V410-Xray3 is well off the
right side of the graph; its extremely red color ($i'-z'=4.374$) is probably due to
the high uncertainty in its $i'$ fit (Section 3.1); since the secondary has an
approximate spectral type of M7.7 (Section 4.3), we expect it to possess colors
similar to those of CFHT-Tau-3: $i'-z'=2$, and thus $i'=17.5$. Also shown are the
average main sequence at the distance of Taurus \citep{hawl02}, 1- and 2-Myr
isochrones based on the evolutionary models of \citet{bcah98}, and a reddening
vector based on the extinction relations reported in Schlegel et al. (1998). The
location of the isochrone is determined by converting the predicted $I_C$
magnitudes to $i'$ magnitudes using the color ($I_C-i'$=0.40) derived as in Section
2 and by assuming the $i'-z'$ colors found from the SDSS field main sequence for
older, more massive dwarfs by \citet{hawl02}. As we show in Section 4.3, the
assumption of dwarf $i'-z'$ colors for these targets appears to be valid. The DUSTY
models (Chabrier et al. 2000) should not be required for these VLMOs due to
their high temperatures at this young age, though we will use these models to
determine limits on faint companions (Section 4.2).

In Figure 4, the binary components are located above the 
SDSS main sequence and well above the background population, so they are most 
likely association members; based on their close proximity and the low 
density of association members we conclude that these are physically 
associated companions. All other detected objects fall well below the empirical 
main sequence and are most likely background stars. One target object, 
KPNO-Tau-12, appears to be somewhat underluminous compared to the other 
Taurus VLMOs. The discovery survey (Luhman et al. 2003a) also found KPNO-Tau-12 
to be underluminous for its assumed age and distance, so this is most likely a 
genuine feature of the system. 

\begin{figure} 
\epsscale{1.00} 
\plotone{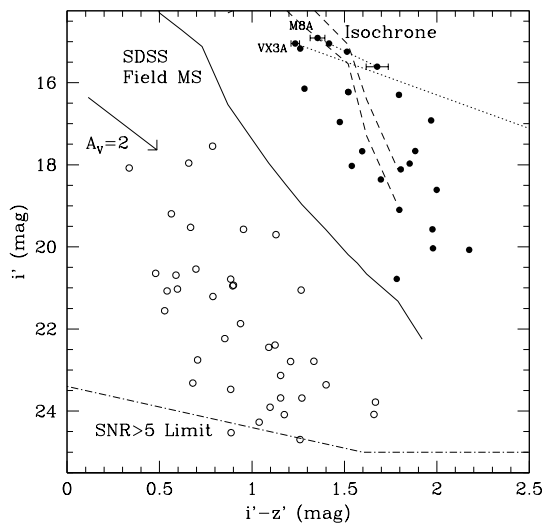} 
\caption{
An $i'$ vs $i'-z'$ color magnitude diagram. Taurus members are shown as filled
circles; objects which are classified as background stars based on their
position in this CMD are shown as open circles.  The SDSS field main sequence
(solid line), 1 and 2 Myr isochrones (dashed lines), and the detection limits of
the survey (dot-dashed line) are also shown. The components of both binary
systems are connected with a dotted line;  the candidate secondary component of
V410-Xray3 is well off the right side of the plot. The $i'$ photometry for
V410-Xray3 B is highly uncertain, so its $i'-z'$ color is not likely to be
accurate. The statistical error bars for binary components are associated with
each point; the error bars for single objects are smaller than the symbols, so
they are not shown.
} 
\end{figure}
\subsection{Sensitivity Limits}

We determined detection limits as a function of distance from the primary stars via
a Monte Carlo simulation similar to that of \citet{met03}. We used the IRAF task
DAOPHOT/ADDSTAR and our average PSF to add artificial stars at a range of radial
separations and magnitudes to the fields of GM Tau, KPNO-Tau-2, KPNO-Tau-7, and
KPNO-Tau-4, which represent the full range of brightness in our sample. We then
attempted to identify the artificial stars with the ALLSTAR PSF-fitting photometry
package. Since the one confirmed VLMO binary in our sample (MHO-Tau-8) has already
been resolved in other observations, our goal for these simulations was to
determine the performance in retrieving known companions. Therefore, unlike similar
simulations described in Kraus et al. (2005), we do not use DAOFIND to identify
sources, but instead report the source positions directly to ALLSTAR for fitting.  
MHO-Tau-8 and V410-Xray3 were the only targets which produced significant fits for
two point sources, so these limits should be appropriate for the detection of new
companions as well.

In Figure 5, we show the detection limits for the four representative targets 
in V and z', as a function of separation, at which we can detect $>10\%$, 
$50\%$, and $90\%$ of the companions. At small separations ($\la$20 AU), the 
50\% detection thresholds roughly scale with the brightness of the primary; 
the survey limits are similar for all objects in terms of $\Delta$$m$. The 
detection limits converge to constant values at large separations; in the case 
where the noise is background-dominated, our $5\sigma$ detection limits are 
$z'$=23.8, $i'=$25.1, and $V$=26.7. The simulations demonstrate that we 
potentially could identify bright, equal-mass pairs as close as 1 pixel 
(0.027\arcsec; 4 AU; $\sim$0.5 $\lambda/D$) and binaries with mass ratio $q=0.1$ 
($\Delta$$z'$=4) at $\ga$10 AU. We also show the locations 
(in $\Delta$$m$ and separation) of the companion to MHO-Tau-8 and the 
candidate companion to V410-Xray3 (for z' only). These results indicate 
that the probability of detecting close binary companions in the z' images 
is near unity for even the faintest systems, but the probability of 
detecting similar systems in the V filter declines rapidly for the 
faintest targets. Our inability to resolve the V410-Xray3 system 
in the V band suggests a limit of $V>20.3$ for the secondary component; the 
corresponding color limit ($V-z'>4.7$) is consistent with the spectral 
type (M7.7) we determine in Section 4.3.

Our simulations suggest that it is not easier to resolve extremely red targets
at shorter wavelengths; given the PSF stability of HST, the improved number
statistics at redder bands outweigh the superior diffraction-limited resolution
at bluer bands. However, the additional color information was useful in
confirming the identify of association members. Also, this is the first large
set of homogeneous observations of young VLMOs shortward of 6000 angstroms. As
we describe in Section 4.3, this allows us to characterize a previously
unexplored regime in their spectral energy distributions.

\begin{figure*}
\epsscale{1.00}
\plotone{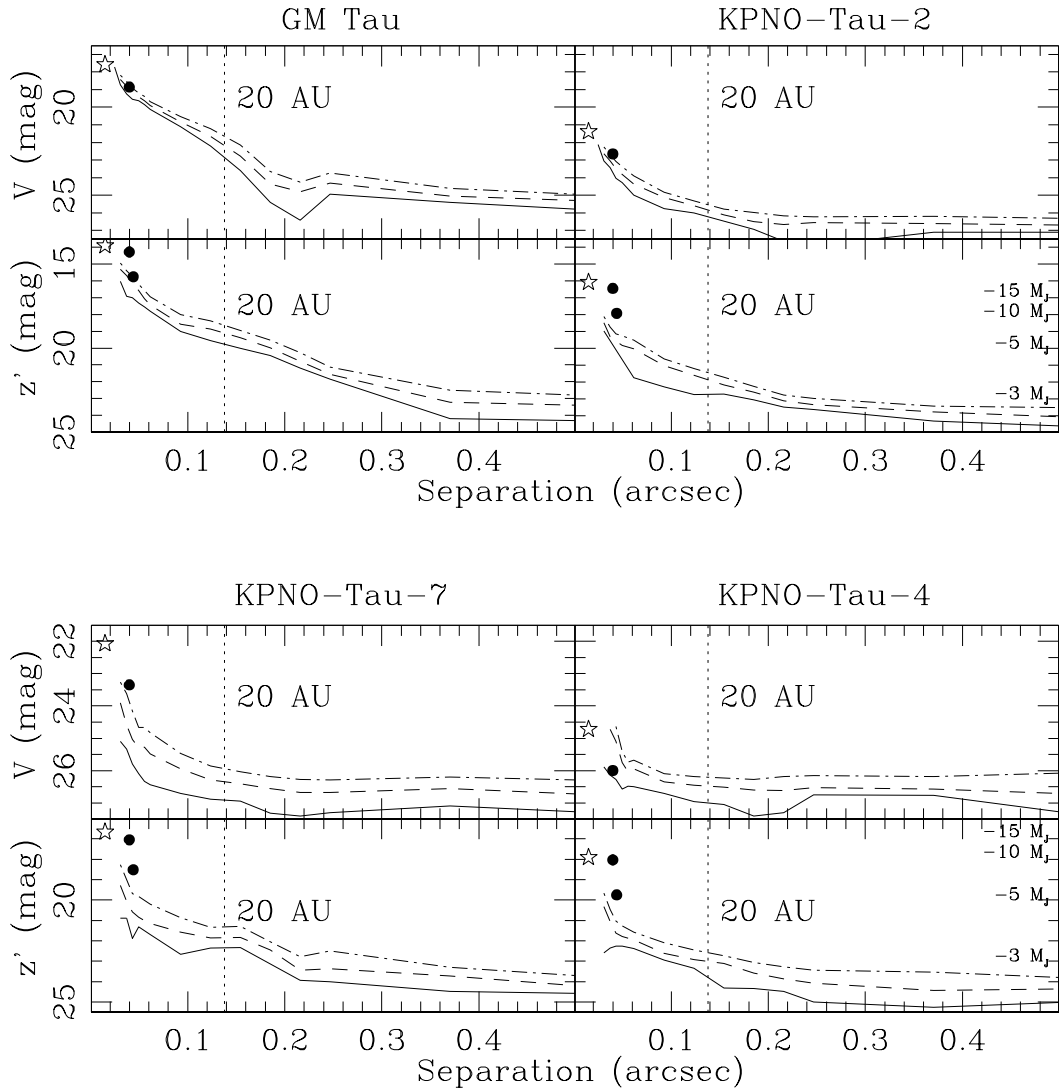}
\caption{
Five-sigma detection frequencies (10\%: solid, 50\%: dashed, and 90\%:
dash-dotted) as a function of separation in each filter for single VLMOs of
maximum, intermediate, and minimum brightness: GM Tau, KPNO-Tau-2 and
KPNO-Tau-7, and KPNO-Tau-4, respectively. Corresponding brightnesses of
potential planetary-mass companions (assuming $A_V=1.5$) are shown on the right
for the z' plots. The brightness of the primary object is denoted with a star to
allow conversion to $\Delta$$m$ values, and the vertical dotted line indicates a
separation of 20 AU at the distance of Taurus (145 pc). The filled circles mark
the separation and $\Delta$$m$ values for the binary companions to MHO-Tau-8 and
V410-Xray3.
}
\end{figure*}
\subsection{Uncertainties in Binary Properties}

A similar Monte Carlo routine was used to test the uncertainties 
in the measurements of MHO-Tau-8 and V410-Xray3. We used ADDSTAR and the 
average PSF to construct 100 simulated images, given the positions and 
brightnesses reported for the primary and the secondary for the real 
images, and then used ALLSTAR to perform PSF-fitting photometry on these 
simulated images. The standard deviation in separation ($\sim$0.5 mas) reported 
for the simulated images of MHO-Tau-8 is consistent with that calculated 
from the standard deviation in separation between the three filters 
($\sim$2 mas), and the standard deviation in the flux ratio $\Delta$$m$ 
(0.04 magnitudes in each filter) is consistent with that determined from 
the magnitudes reported by ALLSTAR for each object. The uncertainties 
predicted by our simulations are generally lower than those observed, but 
this is likely due to the small variations in PSF width observed for each 
target (Section 2.2). We found similar results for the z'-band images of 
V410-Xray3 (0.5 mas, 0.03 magnitudes), but ALLSTAR only found a two-source 
fit for this object in 87\% of the i'-band images, and the standard 
deviation among successful fits was much higher in both separation 
($\sim$7 mas) and flux ratio (0.4 magnitudes). These uncertainties are 
dominated by the measurements of the secondary component, and match the 
uncertainties from the observations. 

We also conducted Monte Carlo tests to determine the probability of 
mistakenly identifying a true single star as a binary. We constructed a 
series of simulated images (100 each for several objects spanning our 
sample's range of brightness), and then tried to fit each object with two 
point sources. We found that this never produced a successful fit, so the 
probability of an erroneous binary identification due to statistical 
errors appears to be low ($<1\%$). However, since the PSF width varies 
between observations, there is also a systematic component to the 
uncertainty in binary values. We have used the average PSF for all fits, 
but since there are no other bright objects in either binary field, it is 
not possible to determine whether our wide or narrow PSFs might be more 
appropriate in a specific case. Experiments show that the separation and 
position angle for each binary are independent from the choice of template 
PSF, but the photometry is not. If we attempt to fit the binaries with the 
wider template PSFs, then ALLSTAR is unable to produce a significant fit. 
If we attempt to fit the binaries with the narrower template PSFs, ALLSTAR 
reports values for $\Delta$$m$ which are systematically lower by 
$\sim$50\%. This suggests that higher-resolution observations will be 
required to determine accurate component fluxes.

\section{Analysis}

\subsection{VLMO Binary Frequency}

For our 22 M5.75-M9.5 targets, we have confirmed a previously-identified binary
VLMO system and identified a new candidate binary, both with separations of
$\sim$6 AU. The observed binary fraction is thus either 4.5$\pm$4.5\% or
9.1$\pm$6.4\% for separations $\ga$4 AU. As can be seen in Figure 5, the
detection of faint companions is difficult at separations comparable to the PSF
width (58 mas in V).  Consequently, the total binary fraction in this separation
range may be higher. However, studies of VLMO binaries in the field and in Upper
Scorpius have found that companions preferentially have mass ratios $q>0.6$
despite the surveys' sensitivity to companions with lower mass ratios (Close et
al. 2003; Bouy et al. 2003; Kraus et al. 2005), which corresponds to
$\Delta$$z<1$ according to the models of Baraffe et al. (1998).  If Taurus VLMOs
have a similar mass ratio distribution, then the detection thresholds in Figure
5 predict that we should identify all VLMO companions at separations $\ga$$4$ AU
in at least $z'$ and likely in the other filters as well.  Even if these limits
are optimistic, the identification of V410-Xray3 as a candidate binary near the
minimum measurable separation, $\sim$5 AU, implies that the survey should be
complete to $\Delta$$z'=2$ (q=0.45) for separations $\ga$4 AU. It is therefore
unlikely that there are many additional binaries at separations $\ga$4 AU with
$q>$0.45 in this sample.

The intrinsically higher luminosity of binaries makes them easier to 
identify in membership surveys, which can also bias the binary frequency 
to larger values. However, MHO-Tau-8 and V410-Xray3 are substantally 
brighter than the detection limits of their discovery surveys and would 
have been easily detected without the additional flux of their companions, 
so this effect should not bias our multiplicity results.

\subsection{Limits on Planetary-Mass Companions}

The high dynamic range of our deep exposures would allow us to 
directly image wide planetary-mass companions. In Figure 5, we indicate the 
predicted brightness of some 
representative masses of planetary companions based on the 1 Myr DUSTY 
models of \citet{cbah00}. These values assume an extinction of 
$A_V$$\sim$1.5, a typical value for objects in Taurus. The extremely red 
colors predicted for planetary-mass objects imply that z' observations 
provide the strictest limits on planetary companions. Based on the lack of 
detections, we conclude that there are no planetary companions with mass 
$\ga$3 $M_J$ at projected separations larger than 280 mas (40 AU) for the 
brightest targets or 140 mas (20 AU) for the faintest targets. For 
comparison, these limits would have allowed for a significant detection of 
the planetary-mass companion to 2MASSWJ 1207334-393254, a substellar 
member of the TW Hya association, which has a projected separation of 
$\sim$54 AU, a flux ratio of $\Delta$$z'$$\sim7$, and a predicted mass of 
$\sim$5 $M_J$ \citep{chauv04}. Other surveys in Cham I (12 targets; Neuhauser et 
al. 2002) and IC 348 (37 targets; Luhman et al. 2005a) which included VLMOs in 
the same mass range ($\la$0.12 $M_{\sun}$) also found no planetary-mass 
companions.

\subsection{Inferred (Sub)stellar Properties}

\begin{deluxetable*}{llccccccc}
\tabletypesize{\scriptsize}
\tablewidth{0pt}
\tablecaption{VLMO (Sub)Stellar Properties\label{tbl2}}
\tablehead{\colhead{Name} & \colhead{SpT\tablenotemark{a}} & 
\colhead{M (M$_\sun$)\tablenotemark{b}} & \colhead{$A_V$} & 
\colhead{$L/L_{MS}$} & \colhead{log L ($L_\sun$)} & \colhead{R ($R_\sun$)} & 
\colhead{Accretor?\tablenotemark{c}} & \colhead{$E(V-V_{MS})$}
}
\startdata
CFHT-Tau-1&M7&0.055&4.8&54&-1.4&0.73&N&-0.59\\
CFHT-Tau-2&M7.5&0.04&2.7&35&-1.6&0.56&N&-0.35\\
CFHT-Tau-3&M7.75&0.035&1.8&29&-1.8&0.50&N&-0.71\\
CFHT-Tau-4&M7&0.055&4.4&210&-0.8&1.45&Y?&-0.61\\
KPNO-Tau-1&M8.5&0.02&1.5&10&-2.3&0.27&N&0.08\\
KPNO-Tau-2&M7.5&0.04&0.7&17&-2.0&0.39&N&-0.25\\
KPNO-Tau-3&M6&0.1&2.0&18&-1.6&0.47&Y&-0.74\\
KPNO-Tau-4&M9.5&0.01&4.0&31&-2.0&0.42&Y?&-1.18\\
KPNO-Tau-5&M7.5&0.04&0.3&51&-1.5&0.68&N&-0.35\\
KPNO-Tau-6&M8.5&0.02&0.7&9&-2.3&0.26&Y&-0.95\\
KPNO-Tau-7&M8.25&0.017&0.1&11&-2.2&0.29&Y&-0.12\\
KPNO-Tau-8&M5.75&0.11&0.7&15&-1.7&0.44&N&-0.02\\
KPNO-Tau-9&M8.5&0.02&1.7&7&-2.5&0.22&N&0.43\\
KPNO-Tau-12&M9&0.014&1.6&3&-2.8&0.14&Y&-2.15\\
KPNO-Tau-14&M6&0.1&4.2&123&-0.8&1.22&N&-0.42\\
MHO-Tau-4&M7&0.055&0.8&126&-1.1&1.12&N&-0.39\\
MHO-Tau-5&M7&0.055&0.3&191&-0.9&1.38&Y?&-0.39\\
MHO-Tau-8 AB&M6+M6.6&0.17&1.7&158&-0.7&1.38&N&-0.38\\
LH 0419+15&M7&0.055&1.2&12&-2.1&0.35&-&-0.22\\
V410 Xray-3 AB&M6+M7.7&0.14&1.5&81&-1.0&0.99&N&-0.54\\
V410 Anon-13&M5.75&0.11&5.6&60&-1.0&0.87&Y&-0.57\\
GM Tau&M6.5&0.08&0.2&49&-1.1&0.74&Y&-1.23\\
USco 55&M5.5&0.010&0.1&18&-1.5&0.49&N&0.18\\
USco 66&M6&0.07&-0.4&16&-1.7&0.44&N&-0.58\\
USco 67&M5.5&0.10&0.4&18&-1.5&0.49&N&0.25\\
USco 75&M6&0.07&0.2&19&-1.6&0.48&N&-0.43\\
USco 100&M7&0.05&-0.2&34&-1.6&0.58&N&-0.40\\
USco 109&M6&0.057&0.4&9.1&-1.9&0.33&N&-0.33\\
USco 112&M5.5&0.10&-0.2&6.8&-1.9&0.30&N&-0.63\\
USco 128&M7&0.05&0.5&10&-2.2&0.32&N&-0.28\\
USco 130&M7.5&0.04&0.4&12&-2.1&0.33&N&-0.28\\
USco 131&M6.5&0.06&1.3&6.7&-2.2&0.27&N&-0.08\\
USco 132&M7&0.05&1.9&14&-2.0&0.37&-&-0.40\\
USco 137&M7&0.05&1.0&3.3&-2.6&0.18&-&-0.28\\
\enddata
\tablenotetext{a}{Spectral types for binary secondaries are inferred from 
photometry presented here; others are from their discovery source (Section 
4.3)}
\tablenotetext{b}{Masses are determined from the models of Baraffe et al. 
(1998).}
\tablenotetext{c}{Muzerolle et al. 2003; Mohanty et al. 2005; Muzerolle et al. 
2005.}
\end{deluxetable*}

\subsubsection{Masses and Spectral Types}

In Table 4, we give the inferred spectral types and masses for all of the 
VLMOs in our sample. Spectral types for single VLMOs and for binary 
primaries in our sample are taken from the discovery sources listed in 
Table 1 and were determined via low- or intermediate-resolution 
spectroscopy. The masses for this sample are estimated from the 2-Myr 
mass-magnitude-temperature relations of \citet{bcah98} and the 
temperature-SpT relations of \citet{luh03b}, and range from 0.015 to 0.14 
$M_\sun$. Large systematic errors may be present in these and all pre-main 
sequence models (e.g. Baraffe et al. 2002; Hillenbrand \& White 2004; Close et al. 
2005; Reiners et al. 2005), so they are best used for relative comparison only. 

The models of Baraffe et al. predict that the mass ratio ($q=m_{s}/m_{p}$)  and
difference in component spectral types ($\Delta$$SpT$) for young VLMO binaries are
a function of the primary-to-secondary flux ratios $\Delta$$m$ with only a minor
mass dependence; we report these quantities as determined from the flux ratio
$\Delta$$z'$ in Table 3 for the detected binary VLMOs, MHO-Tau-8 and V410-Xray3.
The models do not report $z'$ magnitudes so we calculated these from the model
$I_C$ magnitudes, the $I_C-i'$ transformations determined in Section 2, and the
$i'-z'$ colors for the appropriate spectral type. The uncertainties reflect the
uncertainties in the flux ratios and the scatter in this relation in the models,
but do not include any systematic uncertainties from the models. The values
determined from $\Delta$$V$ and $\Delta$$i'$ for MHO-Tau-8 are roughly consistent
with these results, but the values for $\Delta$$z'$ are the only reliable result
for V410-Xray3, so we use $\Delta$$z'$ in all cases for uniformity.

\subsubsection{Luminosities and Extinctions}

Our results, in combination with other surveys, also allow us to determine the
luminosity and extinction for each VLMO by fitting the observed spectral energy
distribution (SED) with a reddened dwarf SED of the same spectral type. This
allows us to test for potential optical excesses that previous observations at
longer wavelengths would not have detected. The VLMO SEDs are constructed from
$Vi'z'JHK$ photometry from this work and from the Two Micron All-Sky Survey
(2MASS; Cutri et al. 2003). Our standard SEDs are constructed from the mean
$i'z'JHK$ photometry for field stars in the SDSS (West et al. 2005) and 2MASS
(Leggett et al. 2002), plus the mean $V-K$ colors reported for the field by Reid
et al. (2003). Typical uncertainties are $\la$0.05 magnitudes for our
observations and $\sim$0.1-0.2 magnitudes for the field star photometry. Any
significant difference between a VLMO SED and the corresponding field SED should
be a result of the intrinsically higher luminosity of our targets (pre-main
sequence VLMOs have larger radii) or extinction due to obscuring material. These
will result in a constant multiplication and a wavelength-dependent
multiplication to the flux in each band, respectively. We solve for these
constants, which correspond to $L/L_{MS}$ and $A_V$, by a least-squares fit of
the set of equations:

\[m_{i,VLMO} = m_{i,field} + \frac{A_i}{A_V}A_V - 2.5 log{\frac{L}{L_{MS}}}\]

where $i$ denotes five filters used in the fit (i'z'JHK) and the reddening 
coefficient for each band is taken from Schlegel et al. (1998). Variations 
in distance will also be included in the luminosity term, but if the depth 
is similar to the apparent width of Taurus-Auriga on the sky ($\sim$10 
deg; 25 pc), these should be relatively minor. 

We list the inferred luminosity ratio, total luminosity, radius, and extinction for
each Taurus VLMO in Table 4. The total luminosity is determined from the K-band
magnitude, corrected for extinction, and the bolometric corrections of Leggett et
al. (2002), which are $BC_K$$\sim$3.1 for mid to late M dwarfs. The radius is
determined from the luminosity ratio ($L/L_{MS} \propto (R/R_{MS})^2$) and the main
sequence mass-radius relations of Baraffe et al.  (1998), which find that
$R_{MS}/R_{\odot} \sim M_{MS}/M_{\odot}$ for M5-M9 dwarfs at ages $>$1 Gyr. We also
performed some experiments in varying the spectral type of the field standard;  
these imply that the uncertainty in spectral type (typically half a subclass)  
corresponds to uncertainties of $\sim$0.5 in $A_V$, $\sim$20\% in $L/L_{MS}$, and
$\sim$10\% in $R$. The uncertainty in the bolometric correction is typically
$\sim$5\% for spectral types later than M5 (Leggett et al. 2002). The statistical
uncertainties from the fitting process are generally insignificant compared to
these systematic uncertainties. These experiments also reveal a degeneracy in the
determination of extinction and luminosity; fitting with an earlier or later
spectral type yields systematically different extinctions and luminosities, but the
goodness of fit does not decrease significantly for small changes ($\la$1
subclass). The same fitting process has been applied to the binaries by considering
the total system brightness in each filter and assigning it the spectral type of
the primary fitting it with a field standard corresponding to the spectral type of
the primary. However, since the observed spectrum is a composite of both objects'
spectra, the results may be biased.

In Figure 6, we present SEDs for the 18 Taurus VLMOs which produced successful fits
to the dwarf SEDs (within $2\sigma$). Four stars could not be fit well and are
discussed below. The solid lines and filled circles denote VLMO SEDs which have
been corrected for extinction and the dashed lines denote the best fit dwarf SEDs
which have been shifted upward by $log(L/L_{MS})$ to correct for their
intrinsically lower luminosity. Typical uncertainties in the SED flux measurements
are $\sim$0.02 dex for the VLMO SEDs, $\sim$0.05 dex for the $r'i'z'JHK$ fluxes in
the dwarf SEDs, and $\sim$0.15 dex for the $V$ band fluxes in the dwarf SEDs. Since
our results typically agree with the dwarf SEDs to within the uncertainties, the
common assumption of dwarf colors for pre-main sequence objects appears to be valid
for this age and mass regime. However, as we discuss in Section 4.3.3, the targets
appear to be systematically brighter than equivalent dwarf SEDs in V, yielding
bluer optical colors. In Figure 7, we present SEDs for the four VLMOs with
significant discrepancies in their fit: KPNO-4, KPNO-6, KPNO-12, and GM Tau. The
first three objects deviate significantly only in their $V$ band measurement, but
the optical and near-infrared portions of the SED for GM Tau appear to be
inconsistent; we can not produce a fit to its dwarf reference by any single
combination of $A_V$ and $L/L_{MS}$. Based on the slope of its NIR SED, GM Tau
appears to have been significantly fainter and/or possessed a significant K-band
excess during its epoch of observation in 2MASS. The excellent agreement between
the optical and near-infrared values for all other objects, which were taken at
different epochs, suggest that they are typically not variable at amplitudes of
$>$0.2 magnitudes.

\begin{figure*}
\epsscale{1.00}
\plotone{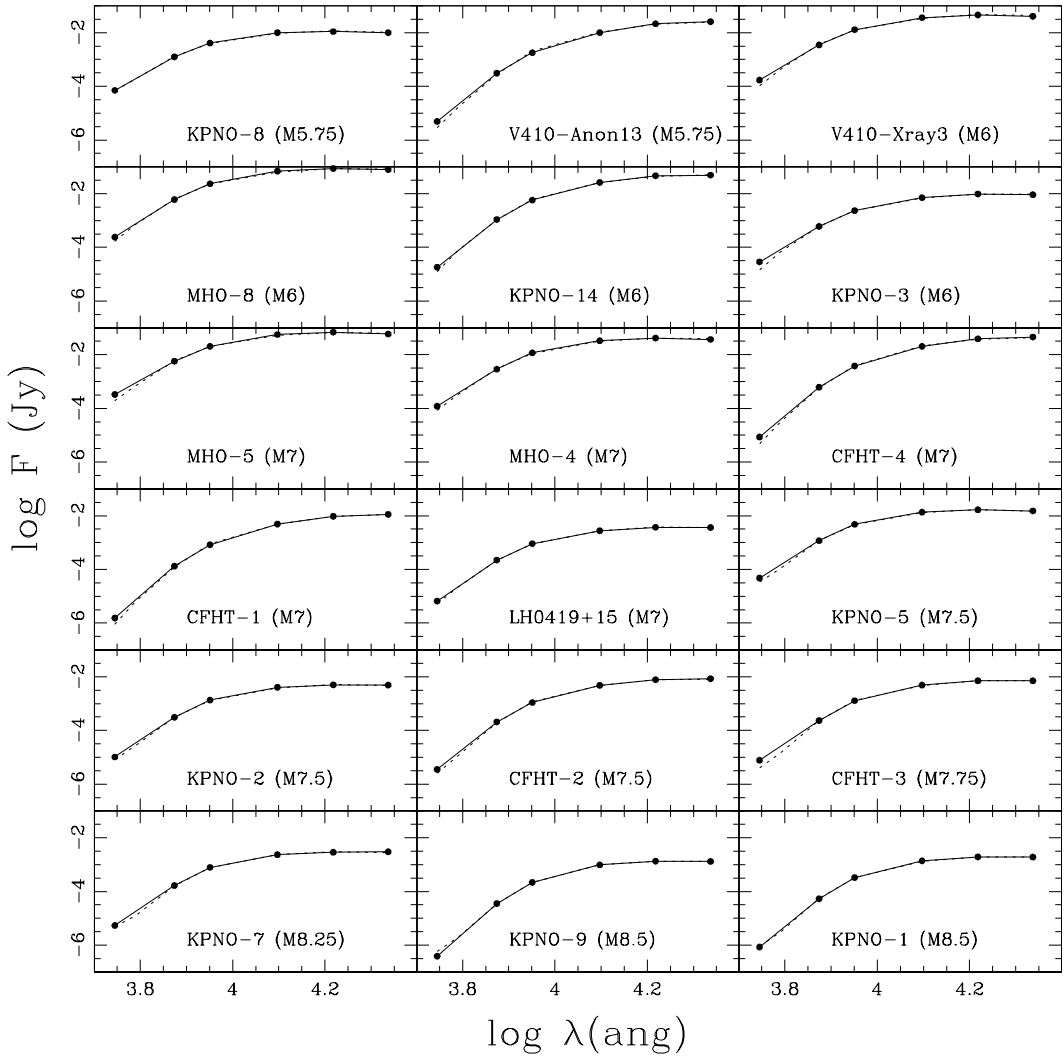}
\caption{
Spectral energy distributions for the VLMOs in our Taurus sample which 
produced successful fits. Solid lines and filled circles denote the 
target SEDs which have been corrected for extinction. Dashed lines denote 
field SEDs which have been corrected for the higher luminosity of our 
targets. 
}
\end{figure*}

\begin{figure*}
\epsscale{1.00}
\plotone{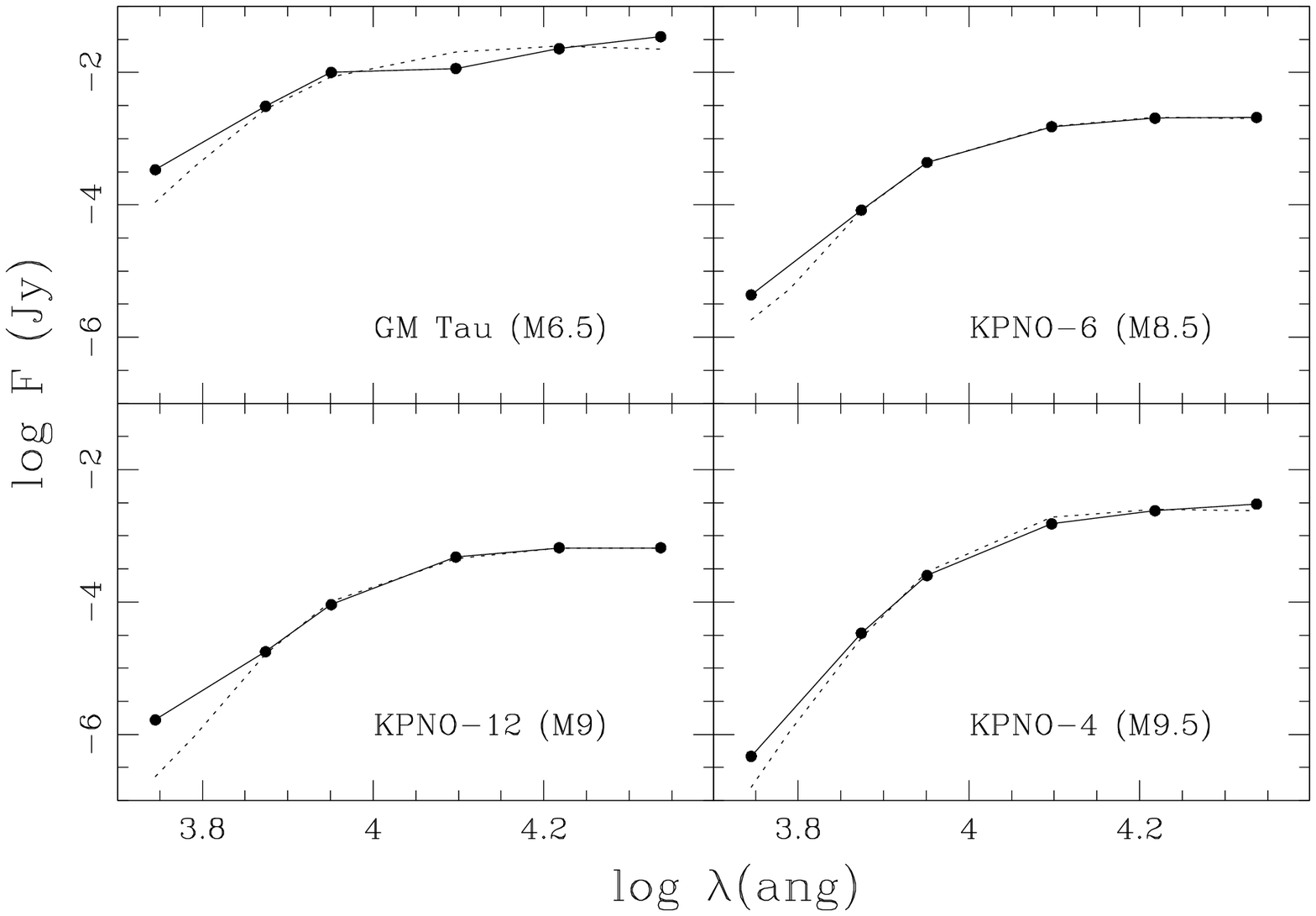}
\caption{Spectral energy distributions for the four VLMOs in our sample 
which produced anomalous fits.
}
\end{figure*}

\subsubsection{Optical Excesses}

Nine of the VLMOs in our sample have been observed to possess spectroscopic 
signatures of active accretion of circumstellar material by Muzerolle et al. 
(2003), Mohanty et al. (2005), and Muzerolle et al. (2005). Accreted material 
falling onto more massive T Tauri stars is typically heated to higher 
temperatures than the stellar photosphere, leading to an optical excess (Basri 
\& Batalha 1990; Hartigan et al. 1991). However, since the rate of accretion 
appears to depend on system mass (e.g. Muzerolle et al. 2003), the low mass 
accretion rates of T Tauri VLMOs should not produce any significant excess. 
Nonetheless, as we note in Section 4.3.2, several VLMOs show signs of an 
optical excess which might be a result of accretion. In Table 4, we identify 
the known accretors and non-accretors and list the V-band excess or deficit (in 
magnitudes) observed for each target. The four objects with the most significant 
V band discrepancies are confirmed or probable accretors.

In Figure 8, we show a plot of the V-band excesses and deficits as a function of
spectral type for each target. Filled circles denote Taurus VLMOs; symbols for
known accretors (based on H$\alpha$ or other line profile diagnostics) are larger
than those of nonaccretors. Open circles denote Upper Sco VLMOs, none of which are
known accretors. The dotted line denotes the V-band excess one would mistakenly
infer if the color were influenced by surface gravity effects, estimated here using
the intrinsically bluer $V-I$ colors of M5-M8 giants (Bessell \& Brett 1988). Most
of the known accretors sit preferentially higher in the plot relative to the
non-accretors and relative to the giant locus, implying that they possess intrinsic
excesses which are related to accretion or to the assessment of their accretion
status.

Even non-accretors seem to possess small V-band excesses over typical dwarf 
colors, though they are not as high as typical giant excesses and do not depend 
significantly on spectral type; this suggests that young VLMOs tend to be bluer 
than dwarfs of the same spectral type. The mean and standard deviation for this 
discrepancy among known non-accretors are 0.28$\pm$0.33 magnitudes; in Figure 8, 
the mean value is denoted with a solid line and the $+2\sigma$ limit with a 
dashed line. If we regard this mean value as a measure of the true photospheric 
flux for young VLMOs, then the four objects identified earlier still possess 
excesses significant at $\ga$$+2\sigma$.

In Table 5, we present the (sub)stellar properties for our Upper Sco sample (Kraus
et al. 2005), which we derive using the methods described in Section 4.3. There are
no known accretors, and the mean V-band excess (0.27$\pm$0.08) is consistent with
the mean excess for nonaccretors in Taurus (0.28$\pm$0.10). As we show in Figure 8,
all Upper Sco VLMOs fall below the $+2\sigma$ criterion for a statistically
significant optical excess over the nonaccretor value. This supports the suggestion
that accretors may show optical excesses over nonaccretors, though the older age of
Upper Sco may also play a role.

\begin{figure*}
\epsscale{1.00}
\plotone{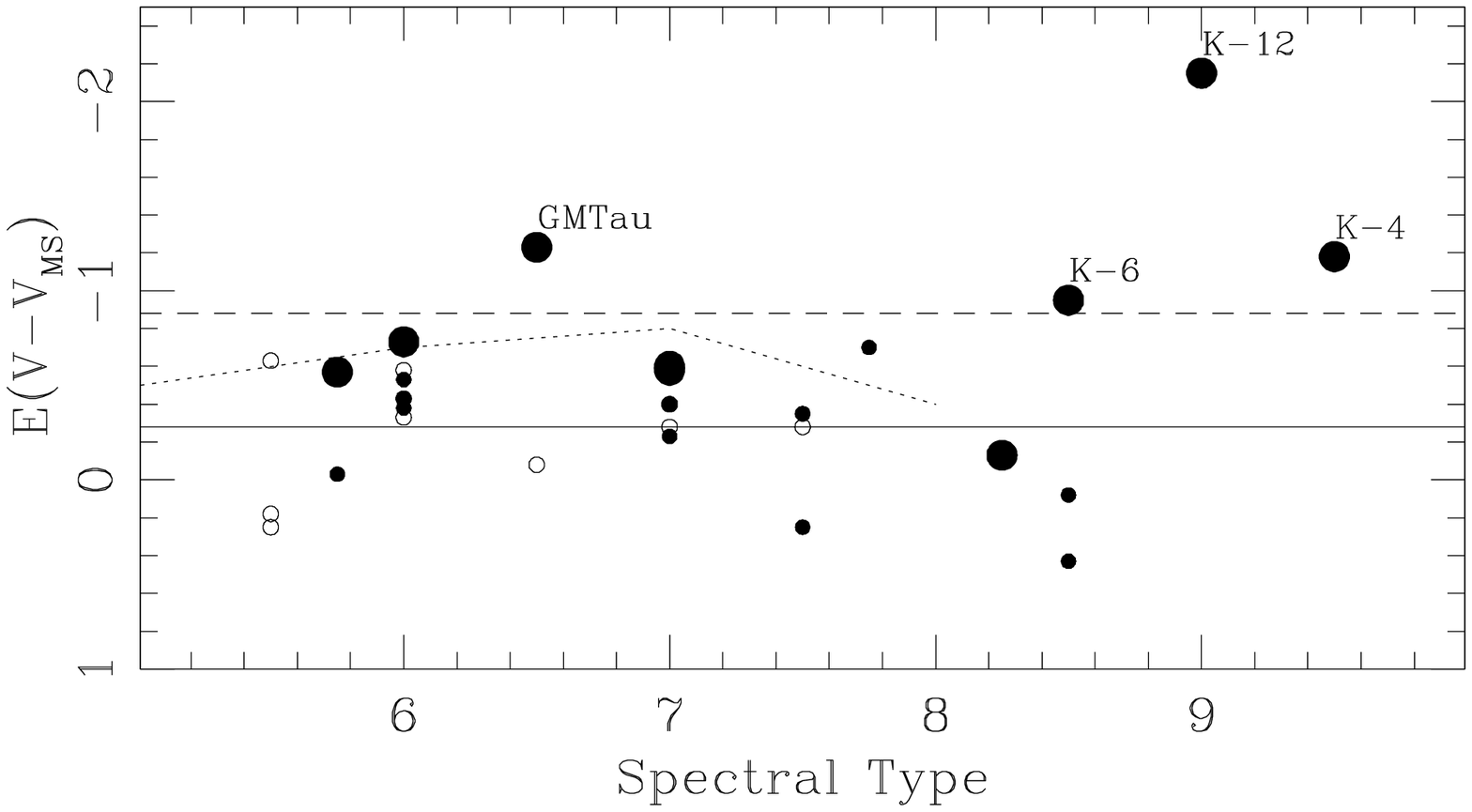}
\caption{V-band excesses as a function of spectral type. Filled circles denote 
Taurus members, open circles denote Upper Sco members, and large 
symbols denote confirmed or probable accretors. A solid line shows the mean value 
for all non-accretors in our sample, and the dashed line shows the corresponding 
$2\sigma$ limit for the existence of significant optical excesses. The four 
objects with significant excesses are labeled.
}
\end{figure*}

\section{Discussion}

\subsection{(Sub)stellar Multiplicity}

Multiplicity surveys of VLMOs have suggested a fairly uniform set of binary
properties in the field \citep{C03,bouy03,burg03,sieg05}, in the Pleiades
\citep{mart03}, and in the young OB Association Upper Scorpius (Kraus et al.
2005). No companions with wide separations ($>$20 AU) or with unequal mass
ratios ($q<0.6$) were found, despite sufficient sensitivity for their detection.
Several wider systems have been discovered serendipitously or during ongoing
surveys (Gizis 2000; Luhman 2004a, 2005c; Phan-Bao et al. 2005; Billeres et al.
2005). However, the corresponding survey statistics are not available, so it is
difficult to determine a frequency for wide VLMO binaries. The binary fractions
observed were also significantly lower than the binary fractions of 57\% for
field G-dwarfs \citep{duq91} and 35-43\% for field M-dwarfs 
\citep{reid97,fisch92}. We summarize these results in Table 5.

\begin{deluxetable}{lccc}
\tabletypesize{\scriptsize}
\tablewidth{0pt}
\tablecaption{Results for Previous Multiplicity Surveys\label{tbl2}}
\tablehead{\colhead{Survey} & \colhead{SpTs} & \colhead{Minimum} & \colhead{Binary} 
\\
\colhead{} & \colhead{} & \colhead{Sep (AU)} & \colhead{Frequency}
}
\startdata
Close et al. (2003)&M8.0-L0.5&3&15$\pm$7\%\\
Bouy et al. (2003)&M8.0-L8.0&1&18.8$\pm$3.7\%\\
Burgasser et al. (2003)&T5.0-T8.0&1&$9^{+4}_{-3}\%$\\
Siegler et al. (2003)&M6.0-M7.5&3&$9^{+4}_{-3}\%$\\
Martin et al. (2003)&M6.0-M9.5\tablenotemark{a}&7&$15^{+15}_{-5}\%$\\
Kraus et al. (2005)&M5.5-M7.5\tablenotemark{a}&5&$25^{+16}_{-9}\%$\\
Current Survey&M5.75-M9.5\tablenotemark{a}&5&$9^{+10}_{-4}\%$\\
\enddata
\tablenotetext{a}{The surveys by Martin et al. and Kraus et al. observed BDs in 
the Pleiades and Upper Sco, respectively. Since these regions are young, 
the spectral type range corresponds to lower masses than in the field.}
\end{deluxetable}

The study presented here is the first large survey of young VLMOs in 
Taurus-Auriga with sufficient resolution to identify binaries at 
separations seen for field VLMOs. Our confirmed binary 
fraction, $4.5\pm4.5\%$ for separations $\ga$4 AU, is lower than, but 
marginally consistent with the results cited for the field. If the binary 
candidate V410-Xray3 is confirmed, the binary fraction in this separation 
range would then be $9.1\pm6.2\%$, which falls within the range of results 
seen in the field. However, the field surveys are sensitive to companions 
with separations as small as $\sim1$ AU and found many companions within 4 
AU that we could not have detected because of the larger distance to our 
targets. Our binary fraction for companions with $a\ga4$ AU only 
sets a lower limit on the total binary fraction for VLMOs. The mass ratio 
and separation for MHO-Tau-8 (q=0.7, a$\sim$6 AU) and the separation 
for V410-Xray3 ($\sim$6 AU) are consistent with the field, but the mass 
ratio for V410-Xray3 may be substantially less. However, this result is 
based only on the z' observations and should be confirmed with followup 
observations.

The methodology of this survey is identical to our Upper Scorpius survey and
their similar distances imply similar spatial resolution, so the two results can
be combined to produce a total VLMO binary fraction (either 4 or 5 of 34,
corresponding to $12\pm6\%$ or $15\pm7\%$) which is also consistent with the
field. Our binary fraction for Taurus (9$\pm$5\%) is somewhat lower than that
for Upper Scorpius (25$\pm$16\%), though the difference is not statistically
significant ($<$2$\sigma$). However, larger surveys (particularly in Upper Sco)
will be required to test whether the binary fraction depends on mass.

\subsection{The Mass Dependence of Multiplicity}

The distinct binary properties of field BDs, relative to those of stars, 
indicate that these properties are mass dependent. However, the form of this 
dependence is not known; \citet{krou03}, \citet{C03}, and Kraus et al. (2005) 
interpret current data as a sharp transition near the stellar/substellar boundary 
while \citet{luh04b} and White et al. (2006) argue for a smooth mass dependence. 
The field binary fraction appears to decrease from 57\% for G dwarfs to 35-43\% 
for M dwarfs (Duquennoy \& Mayor 1991; Fischer \& Marcy 1992; Reid \& Gizis 
1997), but the trend with mass is only marginally significant.

Our combined young VLMO sample for Taurus and Upper Scorpius is large enough to
investigate a possible mass dependence in the VLMO binary frequency over the
single order of magnitude mass range of our survey (0.015-0.12 $M_{\sun}$). All
five binaries and candidate binaries were found among the 13 most massive VLMO
targets ($M\ga0.07$ $M_\sun$); none were found among the 21 lower-mass VLMOs. If
the probability of binarity is 5/34 for all members of our sample, then the
probability that all five binaries will fall among the 13 most massive targets
is (13!/8!)/(34!/29!) $\sim$ 0.005. The possible double-lined spectroscopic
binary KPNO-Tau-14 (Mohanty et al. 2005) also falls in the upper mass range,
though its separation is well inside our survey limit. The detection limits in
Figure 5 demonstrate that bright binaries are more easily identified due to
superior photon statistics, but at least the three high-confidence binaries in
the combined sample (USco-55, USco-66, and MHO-Tau-8) could have been identified
around even the faintest targets.  Also, since this trend is seen at lower
significance in both regions, it should be robust against variations in initial
conditions. The implication is that the binary fraction may decline over the
mass range, though a constant binary fraction combined with a shift toward
smaller separations could produce the same effect in our resolution-limited
sample.

We can further investigate the mass dependence of multiplicity in Taurus by
combining our results with a binary census of more massive Taurus members by White
et al. (2006). This speckle interferometry survey included 44 members of Taurus in
the mass range $0.06<M<1.5$ $M_\sun$ and, in combination with previous results for
41 additional targets (Leinert et al. 1993; Ghez et al. 1993; Simon et al. 1995; 
Duchene 1999), identified 23 binaries with separations $>27$ mas (4 AU).
The sample is reported to be complete for separations of 9-460 AU and mass ratios
$>$0.1 ($\Delta$$K$$<$2.75), which falls within the completeness range of our
survey (Section 3.3). They report that the binary frequency appears to peak for
stars with masses of 0.3-0.7 $M_\sun$ and decline toward lower masses, though the
results are not statistically significant ($<$2$\sigma$). Our VLMO sample supports
this mass dependence, since we found no binaries with separations in the
completeness regime. However, as discussed in Kraus et al. (2005), a similar survey
in Upper Scorpius by Kohler et al. (2000) found no evidence of a similar decline
until immediately above the stellar/substellar boundary (0.1-0.2 $M_{\sun}$). This
implies that the mass dependence of multiplicity could also have a regional
dependence.

\subsection{A Future Dynamical Mass for MHO-Tau-8?}

Theoretical models of the mass-luminosity-temperature relation are still 
largely uncalibrated for low masses and young ages (e.g. Baraffe et al. 
2002). Surface gravity measurements for young brown dwarfs in Upper Sco 
\citep{moh05} have provided some evidence that the theoretical models 
underestimate masses for very low-mass stars and brown dwarfs and overestimate 
masses for the least massive brown dwarfs, but the measurement of dynamical 
masses for low-mass binaries will be required to directly calibrate the models. 
To date, this has been done only for three young VLMOs, the older ($\sim$50-100 
Myr) brown dwarf AB Dor C (Close et al. 2005; Luhman et al. 2005b) and both 
components of the young substellar eclipsing binary 2MASS J05352184-0546085 
(Stassun et al. 2006). 

Relative to the speckle interferometry measurement by White et al. (2006), 
MHO-Tau-8 B has traced $\sim$40 degrees of its orbit around MHO-Tau-8 A within 
only $\sim$5 years; this suggests an orbital period of $\sim$45 years. If we 
assume a circular face-on orbit, then we estimate from Kepler's Law and the 
system parameters ($a\ga6$ AU, $P\sim45$ years) that the total mass of the 
system is $\ga$0.11 $M_\sun$, which is consistent with the total system mass 
implied by the models of Baraffe et al. (1998), $\sim$0.17 $M_\sun$. Since 
this lower limit is substantially lower than the predicted total mass, the 
orbit of the secondary is likely either inclined relative to the plane of the 
sky (implying a larger semi-major axis) or highly eccentric and near apastron. 
In either case, additional astrometric measurements for this system should 
yield precise orbital parameters and better limits on the dynamical masses 
of its components over the next decade.

\subsection{VLMO Optical Excesses}

The rate of (sub)stellar mass accretion onto T Tauri stars is typically estimated 
either by measuring the optical continuum excess or by modeling its 
effect on emission line features that result from outflows as accreting material 
impacts the photosphere. Previous observations of brown dwarfs and low- and 
intermediate-mass stars have found mass accretion rates that scale roughly as $\dot{M} 
\propto M^{2}$ for masses of 0.03-3.0 $M_\sun$ (Gulbring et al. 1998; White \& Ghez 
2001; Muzerolle et al. 2003; Calvet et al. 2004). Specifically, these surveys found 
mass accretion rates ranging from $\sim$$10^{-7}$ M$_\sun$/yr for intermediate-mass T 
Tauri stars with mass 3 M$_\sun$ to $\sim$$10^{-11}$ M$_\sun$/yr for T Tauri-age brown 
dwarfs. Recent results from Mohanty et al. (2005) and Muzerolle et al. (2005), who 
measure accretion rates by modeling of Ca II and H$\alpha$ emission respectively, 
support this trend. They found that many low-mass brown dwarfs (including KPNO-6, 
KPNO-12, and possibly KPNO-4) have apparent mass accretion rates of 
$\sim$5x10$^{-12}$ M$_\sun$/yr. 

These low accretion rates would result in negligible optical excesses
($E(V-V_{MS})<0.01$; e.g. Muzerolle et al. 2003); the very large excesses observed
for our targets imply that either these objects have much higher accretion rates or
the V-band excesses are not the result of accretion. The mass accretion rates that
would be required for each known accretor to generate the optical excess can be
calculated using the method described in Gullbring et al. (1998) whereby the excess
V-band flux is converted to a total accretion luminosity. This luminosity is then
assumed to result from mass infall from the inner edge of a circumstellar disk onto
the surface of the star via magnetospheric accretion. Specifically, we assume that
the radius of the inner disk edge is $5R_{VLMO}$ and the entire change in potential
energy is radiated away via the optical excess. To correct our V-band excess
luminosity to the total accretion luminosity, we adopt a bolometric correction
factor of 10, which is 10\% less than the bolometric correction adopted by White \&
Hillenbrand (2004) for a slightly redder, but narrower band-pass (0.60 - 0.65
$\mu$m).  As emphasized by White \& Hillenbrand, these bolometric corrections are
still highly uncertain, but by adopting a value consistent with previous
assumptions, we can minimize systematic errors in comparisons with previous
calculations. Unfortunately, there are no photometric or spectroscopic data
available at wavelengths shorter than 6000 angstroms for young BDs, so our results 
can not be directly compared to any in the literature.

The resulting mass accretion rates for the 4 stars with significant excesses
($10^{-7}$ to $10^{-9}$ $M_{\odot}/yr$) are 3-4 orders of magnitude higher than the
values found via other methods or inferred from the $M$-$\dot{M}$ relations
observed for higher-mass T Tauri stars. Given this discrepancy, we consider the
possibility that the excess results from our analysis methods or from other
physical processes. However, it is statistically unlikely that 4/22 objects would
possess $>$+2$\sigma$ excesses, or that KPNO-12 would possess a 6$\sigma$ excess,
if these accretors are part of a random distribution with the same mean and
standard deviation as we observe for nonaccretors. Since 4/9 accretors possess
excesses significant at $>$+2$\sigma$ and 8/9 possess excesses significant at
$\ga$+1$\sigma$, it appears that the V-band excesses are related to accretion
diagnostics.  Since three of the outliers fall among the 5 latest objects, there
could also be a gravity-based V-band excess inherent to all late-type young VLMOs.
However, the corresponding trend for giants (which possess even lower surface
gravity) appears to decline for spectral types later than M7 (Figure 8), and there
is no sign of a similar excess for the two non-accreting late-type objects, KPNO-1
and KPNO-9. Finally, the possibility of erroneous spectral classifications cannot
explain this result because there are significant discrepancies in the SED fits for
errors of more than one subclass, and the degeneracy between spectral type, $A_V$,
and $L/L_{MS}$ that we note in Section 4.3.2 acts to preserve most of the inferred
V-band excess under the assumption of a different spectral type.

One explanation that we cannot rule out is the presence of chromospheric activity
or flaring. Stellar flares are characterized by transient regions with higher
temperatures than the surrounding photosphere ($\ga$10$^5$ K); they cause a
temporary increase in the total luminosity, with much higher amplitude of
variability at short wavelengths.  Surveys for x-ray emission from young BDs (e.g.
Preibisch et al.  2005) have found that 1/3 to 1/2 of all objects may possess
significant activity.  These regions also typically produce H$\alpha$ emission; the
width of the base of the H$\alpha$ emission line is a typical test for accretion,
so it is possible that a large flare or significant chromospheric activity could
broaden the line base, mimicking the signature of accretion.

\section{Conclusions}

We present the results of a high-resolution imaging survey of 22 brown dwarfs
and very low mass stars in the nearby ($\sim$145 pc) young ($\sim$1-2 Myr)
low-density star-forming region Taurus-Auriga. This survey confirmed the
binarity of MHO-Tau-8 and discovered a new candidate binary system, V410-Xray3,
resulting in a binary fraction of $9\pm5\%$ at separations $>$4 AU. Both binary
systems are tight ($<$10 AU) and they possess mass ratios of 0.75 and 0.46,
respectively. The binary frequency and separations are consistent with
very-low-mass binary properties in the field, but the mass ratio of V410-Xray3
is among the lowest known. The binary frequency and separations are also
consistent with the trends in multiplicity statistics observed for higher-mass
Taurus members, which suggest a gradual decline in both properties toward low
masses; the implications are that the distinct binary statistics of low-mass
systems are set in the formation process and that this formation process is
similar to that which creates low-mass stars. These objects and a similar set in
Upper Scorpius reveal another possible mass dependence of multiplicity; all five
binaries and candidate binaries fall in the more massive half of the combined
sample, implying either a decline in frequency or a shift to smaller separations
for the lowest-mass binaries. We also combine the survey detection limits with
the models of Chabrier et al. (2000) to show that there are no planets or very
low-mass brown dwarfs with separations $>3 M_J$ at projected separations $>$40
AU orbiting any of the Taurus members in our sample, implying that
planetary-mass companions at wide separations like 2M1207b are found with
frequency $\la$5\%. Finally, we observe significant optical excesses in the
spectral energy distributions of most targets and conclude that the targets with
spectroscopic signatures of accretion possess larger optical excesses than other
young brown dwarfs and very low mass stars.

\acknowledgements

We thank the referee for a prompt, thorough, and helpful critique of this paper.
This work is based on observations made with the NASA/ESA Hubble Space
Telescope, obtained at the Space Telescope Science Institute, which is operated
by the Association of Universities for Research in Astronomy, Inc., under NASA
contract NAS 5-26555. These observations are associated with program \#9853.
This work also makes use of data products from the Two Micron All-Sky Survey,
which is a joint project of the University of Massachusetts and the Infrared
Processing and Analysis Center/California Institute of Technology, funded by the
National Aeronautics and Space Administration and the National Science
Foundation.

\end{document}